\documentclass{llncs}
\usepackage{graphicx}
\usepackage[utf8]{inputenc}
\usepackage[portuguese]{babel}
\usepackage{todonotes}
\usepackage{./why3lang}
\usepackage{hyperref}
\usepackage{pdfpages}

\usepackage{xcolor}
\usepackage{colortbl}   
\usepackage{rotating}

\begin{document}
%

\title{CISE3: Verificação de aplicações com consistência fraca em Why3}

\author{Filipe Meirim  \and
Mário Pereira \and Carla Ferreira}

\institute{NOVA LINCS, FCT, Universidade NOVA de Lisboa, Portugal}
%
%

%
\maketitle              

\pagestyle{plain}
\begin{abstract}

Neste artigo apresentamos uma ferramenta para a verificação formal de programas construídos sobre bases de dados replicadas. Esta ferramenta avalia uma especificação sequencial e deduz quais as operações que necessitam de ser sincronizadas para que o programa funcione correctamente num ambiente distribuído. O nosso protótipo é construído sobre a plataforma de verificação dedutiva Why3. A \textit{framework} do Why3 proporciona uma experiência de utilização sofisticada, a possibilidade de escalar para casos de estudo realistas, assim como um elevado grau de automação de provas. Um caso de estudo é apresentado e discutido, com o propósito de validar experimentalmente a nossa abordagem.
\end{abstract}
\section{Introdução}

Atualmente as aplicações distribuídas de larga escala dependem de sistemas de armazenamento~geo-replicados para melhorarem a experiência do utilizador. Estes sistemas de armazenamento mantêm várias réplicas dispersas geograficamente, que guardam cópias dos dados da aplicação. A geo-replicação permite que haja uma menor latência nas operações e maior disponibilidade da aplicação, pois os pedidos dos clientes são encaminhados para a réplica mais próxima. No entanto a geo-replicação apresenta um potencial problema em relação à integridade da informação armazenada em diferentes réplicas. Quando ocorrem atualizações concorrentes sobre diferentes réplicas, a integridade dos dados pode ser comprometida. Para endereçar esta situação, uma solução é a introdução de sincronização no sistema. Caso o sistema opte por um modelo de consistência forte, em que é garantida uma ordem total na execução de operações nas diferentes 
réplicas, a integridade dos dados é garantida, mas há um decréscimo na disponibilidade do sistema.
Se o sistema usar um modelo de consistência  fraco, que não garante uma ordem total na execução de operações, este fica mais disponível para o utilizador, no entanto, há a possibilidade de perda da integridade dos dados da aplicação. 
Para se encontrar um equilíbrio entre a correção e a disponibilidade, foi proposto que os sistemas geo-replicados poderiam combinar modelos de consistência forte e fraca~\cite{balegas2015,balegas2016}. 
Mais especificamente, a abordagem consiste em usar consistência  forte quando a correção da aplicação está em risco, e beneficiar da consistência  fraca quando as operações são seguras. 
Obter o equilíbrio  entre consistência  fraca e  forte de forma a maximizar a disponibilidade do sistema não é uma tarefa trivial \cite{CAP}. Os programadores necessitam de ponderar sobre os efeitos concorrentes das operações, e de seguida decidir quais destas necessitam de sincronização para garantir a correção do sistema.

Neste artigo, propomos uma abordagem automática para a verificação de aplicações distribuídas com consistência fraca utilizando a plataforma de verificação dedutiva Why3 \cite{filliatre:hal-00789533}. Esta abordagem segue a regra de prova proposta pela lógica CISE \cite{cise}. O objetivo desta regra de prova é identificar quais os pares de operações que não podem ser executados em concorrência, e que por consequência necessitam de sincronização. Neste artigo, apresentamos a nossa abordagem através de exemplos escritos e verificados na nossa ferramenta.

A Secção \ref{cise} apresenta uma breve introdução à lógica CISE e à regra de prova subjacente. A plataforma Why3 é apresentada na Secção \ref{why3}. Na Secção \ref{cise3} apresentamos a nossa abordagem e demonstramos o seu funcionamento. Na Secção \ref{case_study}, ilustramos o funcionamento da nossa ferramenta através dum caso de estudo completo. Na Secção \ref{related_work} exibimos algumas ferramentas semelhantes à nossa. Concluímos na Secção \ref{conclusion}, apresentando linhas de exploração futura.

\section{Conceitos preliminares} \label{cise}

A lógica CISE \cite{cise} apresenta uma regra de prova para verificar a preservação de invariantes de integridade de aplicações que operam sobre bases de dados replicadas. Esta lógica fornece um modelo de consistência genérico que permite definir modelos de consistência especializados para as operações do sistema. 
Um modelo de consistência é expresso através de um sistema de \emph{tokens}, contendo um conjunto dos \emph{tokens} $T$ e uma relação de conflito $\bowtie$ sobre $T$.
Cada operação do sistema tem associado um conjunto de \emph{tokens} que deve adquirir para poder ser executada.
Operações com \emph{tokens} em conflito não podem ser executadas em concorrência e necessitam de ser sincronizadas. Contudo, se o conjunto de \emph{tokens} associado a uma operação é vazio, então essa operação não necessita de ser sincronizada e pode ser usada consistência fraca. 

Por exemplo, o protocolo de exclusão mútua pode ser expresso assumindo um único \emph{token} $\tau$ em conflito consigo próprio ($\tau \bowtie \tau$). 
Sendo que as operações que acedem ao recurso partilhado devem estar associadas ao \emph{token} $\tau$.
Assim, é garantido que quando uma operação adquire o token $\tau$, nenhuma outra operação  que necessite desse token poderá ser executada em concorrência. 
Ou seja, é garantida exclusão mútua.

A regra de prova proposta permite duas abordagens de verificação. A primeira abordagem consiste em verificar a validade de um sistema de \emph{tokens} previamente definido
pela programadora.
A segunda abordagem consiste em identificar os pares de operações (em conflito) que quando executadas em concorrência podem quebrar os invariantes da aplicação.
Estes pares de operações definem já um sistema de \emph{tokens} inicial, que pode ser posteriormente refinado pela programadora.  

Em termos operacionais, a regra de prova é composta pelas seguintes análises.
\textbf{Análise de \textit{safety}:} é verificado se os efeitos de uma operação, quando executada sequencialmente, preservam o invariante da aplicação.
\textbf{Análise de comutatividade:} é verificado se todos os pares de operações comutam, i.e., se as operações forem executadas em ordem alternada então o estado final será o mesmo, partindo do mesmo estado inicial.
\textbf{Análise de estabilidade:} é verificado se as pré-condições de uma operação são estáveis em relação aos efeitos de cada uma das operações do sistema.
A estabilidade de uma operação garante que esta pode ser executada em concorrência com outras operações de forma segura, i.e., aquando da execução remota as suas pré-condições continuam válidas.

A regra de prova foi automatizada numa ferramenta com o propósito de raciocinar sobre a correção de aplicações distribuídas, executadas sobre bases de dados com consistência fraca. A automatização da regra de prova usa \textit{SMT solvers} para descartar condições de verificação geradas.

Na primeira versão da ferramenta \cite{Najafzadeh:2016:CTP:2911151.2911160} é necessário comunicar diretamente com o Z3, o que torna a utilização da ferramenta complexa, pois é necessário utilizar API's de baixo nível para especificar a aplicação~\cite{nair:hal-01832888}. Na segunda versão da ferramenta \cite{Marcelino:2017:BHC:3064889.3064896} é usada a \textit{framework} Boogie \cite{boogie}. O Boogie gera um conjunto de condições de verificação a partir da especificação, e de seguida envia-as a um \textit{SMT solver}. Contudo, uma desvantagem de usar o Boogie é a falta de suporte~atual.

\section{Plataforma Why3} \label{why3}

Why3 é uma plataforma de verificação dedutiva de programas, isto é, "torna a correção dum programa numa fórmula matemática e de seguida prova-a" \cite{memoire}. A arquitetura do Why3 contém um \textit{back-end} puramente lógico e um \textit{front-end} direcionado para a escrita de programas \cite{memoire}. O \textit{front-end} recebe ficheiros que contêm uma lista de módulos, dos quais condições de verificação são extraídas, a serem posteriormente enviadas a demonstradores de teoremas externos.

Esta ferramenta oferece uma linguagem de programação denominada WhyML que tem dois propósitos: escrita de programas e fornecer especificações formais dos seus comportamentos. A linguagem WhyML, constitui-se como uma linguagem de primeira ordem com alguns aspetos encontrados frequentemente em linguagens funcionais, como por exemplo, \textit{pattern-matching}, tipos algébricos e polimorfismo. Em simultâneo, a linguagem WhyML apresenta características imperativas como \textit{records} com campos mutáveis e exceções. A lógica usada para escrever as especificações é uma extensão da lógica de primeira ordem com tipos polimórficos, tipos algébricos, predicados indutivos, definições recursivas, assim como uma forma limitada de lógica de ordem superior \cite{modular_way}. Outra característica importante da linguagem WhyML é código \textit{ghost} utilizado principalmente quando se escreve a especificação do programa. Este tipo de código é semelhante ao código regular pois é \textit{parsed}, \textit{type checked} e transformado em condições de verificação da mesma forma \cite{memoire}. Uma particularidade do código \textit{ghost} é o facto de poder ser removido dum programa sem afetar o seu funcionamento. A frase anterior é verdade pois o código \textit{ghost} não pode ser usado numa computação de código regular, assim como também não pode alterar um valor mutável no código regular \cite{ghost_code}. Mas código regular não pode modificar nem aceder a código \textit{ghost}. A linguagem WhyML também pode ser usada como uma linguagem intermédia para a verificação de programas escritos em C, Ada ou Java, por exemplo \cite{filliatre:hal-00789533,Kirchner:2015:FSA:2769048.2769092,spark,krakatoa}. A plataforma Why3 fornece
um \textit{front-end} para comunicar com mais de 25 demonstradores de teoremas automáticos e interativos. 

\section{CISE3} \label{cise3}

Como mencionado na secção anterior, a plataforma Why3 fornece uma linguagem de programação e especificação de alto nível. A partir de programas escritos e anotados em WhyML são posteriormente geradas condições de verificação, que podem ser enviadas aos múltiplos demonstradores de teoremas integrados no Why3. 
Esta diversidade permite colmatar algumas limitações de outras ferramentas de prova que apenas suportam um único demonstrador de teoremas. 
Em simultâneo, a plataforma Why3 também fornece um ambiente de desenvolvimento gráfico onde é possível interagir com os diversos demonstradores de teoremas, realizar
provas interativas~\cite{dailler2018}, assim como visualizar contra-exemplos.

A plataforma Why3 pode ser estendida através de \textit{plug-ins} como é o caso do Jessie \cite{jessie} e o Krakatoa \cite{krakatoa}. A integração de novos \textit{plug-ins} no Why3 é  simples: 
é necessário escrever um \textit{parser} para a  linguagem alvo, cuja representação intermédia deve ser mapeada para a \textit{AST} não tipificada da linguagem WhyML. Finalmente, 
deve ser utilizado o mecanismo de tipificação do Why3 para gerar uma versão tipificada da anterior \textit{AST}. O programa WhyML resultante da tipificação pode então ser analisado de forma transparente pela plataforma.

A ferramenta CISE3 trata-se de um \textit{plug-in} da plataforma Why3. No nosso caso particular, utilizamos o \textit{parser} já existente para a linguagem WhyML, assim como o \textit{printer} de expressões. Acreditamos que a nossa escolha da plataforma Why3, baseando a nossa ferramenta na sua arquitetura de \textit{plug-ins}, permitiu uma redução do esforço de desenvolvimento e validação. Por outro lado, construir a nossa ferramenta sobre uma \textit{framework} matura permite-nos evoluir para a análise de exemplos mais realistas. Nesta secção descrevemos o funcionamento da ferramenta CISE3 e ilustramos as várias análises realizadas.

\paragraph{Análise de \textit{safety}.} \label{safety}
Esta análise consiste em verificar se uma operação executada sem concorrência pode quebrar algum invariante de integridade da aplicação. O \textit{input} necessário à nossa ferramenta é a especificação do estado da aplicação e seus invariantes, assim como a implementação e especificação para cada operação. Consideremos, por exemplo, o seguinte programa genérico constituído pelas operações \of{f} e \of{g}, assim como o tipo de estado $\tau$ e o invariante \of{I} associado:
\small{\begin{why3}
  type $\tau$[@state] = { $\overline{\mathtt{x:\tau_x}}$ }
  invariant { I }

  let f ($\overline{\mathtt{x:\tau_1}}$) (state: $\tau$)
    requires { $\mathcal{P}_1$ }
    ensures  { $\mathcal{Q}_1$ }
  = e$_1$

  let g ($\overline{\mathtt{y:\tau_2}}$) (state: $\tau$)
    requires { $\mathcal{P}_2$ }
    ensures  { $\mathcal{Q}_2$ }
  = e$_2$
\end{why3}}

Para especificar o estado da aplicação, é necessário associar a esse tipo uma etiqueta \of{[@state]} de forma a ser possível identificá-lo, como podemos observar acima. O tipo~$\tau$ possui um conjunto de campos representado por $\overline{\texttt{x}}$ cujos tipos são representados por $\overline{\tau_{\texttt{x}}}$. Adicionalmente, apresentamos a definição das operações \of{f} e \of{g}. Todas as operações da aplicação têm como argumento uma instância do estado da aplicação para ser usado na geração de funções para as análises de estabilidade e comutatividade. No código acima, na operação \of{f} as suas pré-condições são representadas por $\mathcal{P}_1$, as pós-condições por $\mathcal{Q}_1$ e o seu corpo está representado por \of{e}$_1$. A especificação da operação \of{g} é semelhante mas as suas pré-condições são representadas por $\mathcal{P}_2$, as pós-condições por $\mathcal{Q}_2$ e o seu corpo por \of{e}$_2$.
A plataforma Why3 por si só já tem a capacidade de verificar, operação a operação, se as suas pré e pós-condições são mantidas tendo em conta a sua implementação. Em adição, a programadora ao especificar o estado da aplicação juntamente com os seus invariantes de integridade, o Why3 consegue verificar se uma operação os pode quebrar ou não, dada a sua implementação. Nos casos em que o Why3 não consiga provar alguma asserção no programa, a plataforma é capaz de apresentar um contra-exemplo~\cite{DBLP:journals/jlp/DaillerHMM18}. Dito isto, é possível afirmar que o Why3 é capaz de realizar a análise de \textit{safety}, sem ser necessário realizar qualquer alteração interna à plataforma. 

\paragraph{Análise de comutatividade e estabilidade.} \label{comm}
O nosso \textit{plug-in} foi implementado com o intuito de gerar automaticamente funções que verificam a comutatividade e estabilidade entre pares de operações da aplicação. Este \textit{plug-in} dado o código e especificação da aplicação usa o \textit{parser} do Why3 para obter uma representação em memória do conteúdo de um programa WhyML. Após o \textit{parsing} do ficheiro, para cada par de operações diferentes é gerada automaticamente uma função de comutatividade. Voltando ao programa genérico apresentado na secção anterior, a nossa ferramenta gera a seguinte função para a análise de comutatividade e estabilidade entre as operações~\of{f}~e~\of{g}:
\begin{why3}
  let ghost f_g_commutativity () : ($\tau$, $\tau$)
    returns { (s$_1$, s$_2$) $\rightarrow$ s$_1$ == s$_2$ }
  = val $\overline{\mathtt{x_1: \tau_1}}$ in
    val state$_1$ : $\tau$ in
    val $\overline{\mathtt{x_2: \tau_2}}$ in
    val state$_2$ : $\tau$ in
    assume { $\mathcal{P}_1 \wedge \mathcal{P}_2 \; \wedge $ state$_1$ == state$_2$ }
    f $\overline{\mathtt{x}_1}$ state$_1$;
    g $\overline{\mathtt{x}_2}$ state$_1$;
    g $\overline{\mathtt{x}_2}$ state$_2$;
    f $\overline{\mathtt{x}_1}$ state$_2$;
    (state$_1$, state$_2$) 
\end{why3}

Se o Why3 conseguir provar que a função acima respeita a especificação, então as operações comutam e não estão em conflito. Esta função começa por gerar dois estados da aplicação iguais assim como os argumentos para cada operação de modo a que se preservem as pré-condições de ambas as operações em análise. De seguida ambas as operações são executadas numa ordem sobre um dos estados gerados e depois são executadas na ordem alternativa sobre o outro estado gerado. No final é verificado se os estados que são obtidos após a execução de ambas as operações em ordens alternadas são iguais. Caso ambos os estados finais sejam iguais, então assumimos que as operações são comutativas. Para verificar que as operações não estão em conflito, tenta-se provar a função gerada com a nossa ferramenta. Caso não consigamos provar uma pré-condição de uma operação que tenha sido precedida pela execução de outra operação (\textit{e.g.,} a pré-condição de \of{g} no estado obtido após a execução de \of{f}), então assume-se que estas estão em conflito e não podem ser executadas em concorrência. Caso contrário as operações não estão em conflito. Desta forma consegue realizar-se num único passo a análise de comutatividade e estabilidade para todos os pares de operações diferentes.

Como se pode observar na pós-condição da função acima, existe uma relação de igualdade de estados (\of{==}). Esta relação de igualdade trata-se de uma comparação campo a campo do \textit{record} do estado da aplicação. A relação de igualdade entre estados pode ser gerada automaticamente pela nossa ferramenta. Na Secção \ref{case_study} apresentaremos um exemplo que requer uma relação de igualdade fornecida pela programadora.

Finalmente, para a análise de estabilidade para cada par de operações iguais, a nossa ferramenta também gera uma função. Voltando, mais uma vez, ao programa genérico apresentado na secção anterior, a nossa ferramenta gera as seguintes funções para a análise de estabilidade para as operações \of{f} e \of{g}:

\begin{why3}
  let ghost f_stability () : unit
  = val $\overline{\mathtt{x_1: \tau_1}}$ in
    val state$_1$ : $\tau$ in
    assume { $\mathcal{P}_1 $}
    f $\overline{\mathtt{x}_1}$ state$_1$;
    f $\overline{\mathtt{x}_1}$ state$_1$;
  
  let ghost g_stability () : unit
  = val $\overline{\mathtt{x_2: \tau_2}}$ in
    val state$_1$ : $\tau$ in
    assume { $\mathcal{P}_2 $}
    g $\overline{\mathtt{x}_2}$ state$_1$;
    g $\overline{\mathtt{x}_2}$ state$_1$;
\end{why3}

Nas funções acima apresentadas, começa-se por gerar o estado inicial da aplicação e os argumentos para as chamadas da operação de forma a que as pré-condições da primeira chamada à operação sejam preservadas. Por último a operação é executada duas vezes consecutivas. Para determinar se a operação é independente de si própria, utiliza-se o Why3 para se tentar provar essa função. Caso não se consiga provar a função, devido à violação de uma pré-condição da segunda chamada à operação, então assume-se que a operação não pode ser executada em concorrência consigo própria.

\section{Caso de estudo} \label{case_study}

Nesta secção apresentamos um caso de estudo completo, que visa ilustrar o funcionamento da nossa ferramenta. O caso de estudo trata-se de um conjunto de contas bancárias, sobre as quais cada cliente pode apenas depositar ou levantar dinheiro. A implementação e especificação são as seguintes:

\begin{why3}
  type state [@state]= {
    balance : array int
  } invariant{forall i. 0 <= i < length balance -> balance[i] >= 0}

  let deposit(accountId amount: int) (state : state): unit
    requires {amount > 0}
    requires {accountId >= 0 /\ accountId < length state.balance}
    ensures  {state.balance[accountId] = 
              old(state.balance)[accountId] + amount}
    ensures  {forall i. i <> accountId -> 
              state.balance[i] = (old state.balance)[i]}
  = state.balance[accountId] <- state.balance[accountId] + amount

  let withdraw(accountId amount: int) (state : state) : unit
    requires {amount > 0}
    requires {state.balance[accountId] - amount >= 0}
    requires {accountId >= 0 /\ accountId < length state.balance}
    ensures  {state.balance[accountId] = 
              old(state.balance)[accountId] - amount}
    ensures  {forall i. i <> accountId -> 
              state.balance[i] = (old state.balance)[i]}
  = state.balance[accountId] <- state.balance[accountId] - amount

  let ghost predicate state_equality [@state_eq] (s1 s2 : state)
  = array_eq s1.balance s2.balance
\end{why3}

\begin{figure}
    \centering
    \includegraphics[scale=.75]{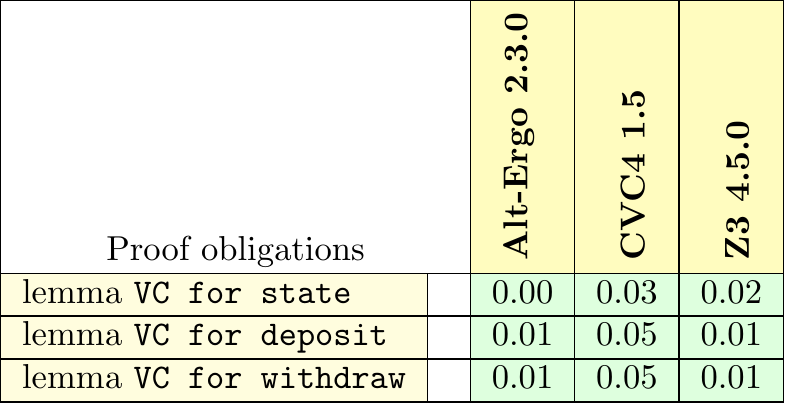}
    \vspace{-5pt}
    \caption{Estatísticas da prova da aplicação do caso de estudo.}
    \label{fig:table1}
\end{figure}

A especificação das operações \of{deposit} e \of{withdraw} são standard: não é possível depositar valores negativos nem é possível remover uma quantia de dinheiro maior que o saldo disponível. O estado da aplicação especificado corresponde a um \textit{record} com apenas um campo \texttt{balance} que representa as contas bancárias, e cada índice desse \texttt{array} representa uma conta diferente. Ao estado da aplicação está associado um invariante de integridade. Para este exemplo, o invariante especifica que, em qualquer momento da execução, o saldo de cada conta deve ser não-negativo. O esforço de prova deste programa sequencial é apresentado na Figura \ref{fig:table1}.

Uma particularidade deste exemplo é a introdução do predicado de igualdade entre estados \of{state_equality}. Este predicado está identificado como sendo código \textit{ghost} pois é apenas utilizado para propósitos de especificação. De forma a identificarmos que o predicado \of{state_equality} é a relação de igualdade entres estados, é necessário associar-lhe a etiqueta \of{[@state_eq]}. Caso a programadora não tivesse especificado este predicado, a nossa ferramenta teria gerado automaticamente uma relação de igualdade, como vimos na Secção \ref{cise3}. No caso concreto da comparação entre \textit{arrays}, uma simples comparação estrutural não seria suficiente para provar a igualdade entre estes, já que é necessário fazer uma comparação \textit{point-wise} dos elementos dos \textit{arrays}.

Na implementação acima apresentada podemos observar que a aplicação apenas tem duas operações: \of{deposit} e \of{withdraw}. Neste caso havendo apenas um par de operações diferentes a nossa ferramenta apenas necessita de gerar a seguinte função para a análise de comutatividade:

\begin{why3}
  let ghost deposit_withdraw_commutativity () : (state, state)
    ensures  { match result with
               | x1, x2 -> state_equality x1 x2
               end } 
=   val ghost accountId1 : int in
    val ghost amount1 : int in
    val ghost state1 : state in
    val ghost accountId2 : int in
    val ghost amount2 : int in
    val ghost state2 : state in
    assume { ((amount1 > 0 /\ accountId1 >= 0 /\ 
               accountId1 < length (balances state1)) /\
               amount2 > 0 /\
             ((balances state2)[accountId2] - amount2) >= 0 /\
               accountId2 >= 0 /\ 
               accountId2 < length (balances state2)) /\
               state_equality state1 state2 };
    withdraw accountId2 amount2 state1;
    deposit accountId1 amount1 state1;
    deposit accountId1 amount1 state2;
    withdraw accountId2 amount2 state2;
    (state1, state2)

\end{why3}

Na função acima começa-se por gerar os argumentos \of{amount1}, \of{accountId1}, \of{state1}, \of{amount2}, \of{accountId2} e \of{state2} que serão usados nas chamadas às operações em análise. A expressão \of{assume} tem o propósito de limitar o espaço de possíveis combinações de valores para os argumentos que são gerados, de modo a que as pré-condições sejam preservadas e ambos os estados iniciais gerados sejam iguais. Posteriormente é chamada a operação \of{withdraw} seguida da chamada à operação \of{deposit} sobre o estado \of{state1}. Se o Why3 não conseguir provar as pré-condições para a chamada da operação \of{deposit} sobre o estado \of{state1}, então assume-se que as operações estão em conflito e não podem ser executadas em concorrência. De seguida é chamada a operação \of{deposit}, seguida da chamada à operação \of{withdraw} sobre o estado \of{state2}. Da mesma forma que foi verificado na ordem de execução anterior, se alguma pré-condição da chamada à operação \of{withdraw} sobre o estado \of{state2} não for preservada então assume-se que as operações estão em conflito. Para finalizar, é devolvido um par com os estados \of{state1} e \of{state2} e se forem ambos iguais então assume-se que as operações comutam. Neste caso em particular conseguimos provar todas as condições de verificação geradas, nomeadamente, a preservação das pré-condições das operações e a igualdade dos estados após a execução das operações em ambas as ordens. Concluímos assim que as operações \of{deposit} e \of{withdraw} comutam e não estão em conflito, sendo possível executá-las em concorrência.

Procede-se agora à análise de estabilidade para os pares de operações iguais. Nesses casos, gera-se automaticamente uma nova função de análise de estabilidade para cada par de operações iguais. Para o nosso caso de estudo, a nossa ferramenta gera as seguintes funções para a restante análise de estabilidade:

\begin{why3}
  let ghost deposit_stability (ghost _:()) : () =
    val ghost accountId : int in
    val ghost amount : int in
    val ghost state : state in
    assume { amount > 0 /\ 
             accountId >= 0 /\ 
             accountId < length (balances state) };
    deposit accountId amount state;
    deposit accountId amount state
\end{why3}
\begin{why3}
  let ghost withdraw_stability (ghost _:()) : () =
    val ghost accountId : int in
    val ghost amount : int in
    val ghost state : state in
    assume { amount > 0 /\ 
             ((balances state)[accountId] - amount) >= 0 /\
             accountId >= 0 /\ accountId < length (balances state) };
    withdraw accountId amount state;
    withdraw accountId amount state
\end{why3}

Inicialmente cada função de análise de estabilidade gera os argumentos para serem usados nas chamadas às funções, neste caso \of{amount}, \of{state} e \of{accountId}. A expressão \of{assume} 
tem o propósito de restringir o espaço de combinação de valores que os argumentos podem tomar de forma a que as pré-condições da operação em análise sejam preservadas. De seguida chama-se a operação duas vezes consecutivas e caso o Why3 não consiga provar a preservação de todas as pré-condições \textit{a priori} da segunda chamada à operação em análise então assume-se que a operação está em conflito consigo própria. Para este caso de estudo, todas as condições de verificação geradas para a função \of{deposit_stability} são provadas automaticamente, o que nos permite concluir que a operação \of{deposit} pode ser executada em concorrência consigo própria. Por outro lado, não é possível provar a preservação de uma das pré-condições aquando da segunda chamada à operação \of{withdraw} na função \of{withdraw_stability}. Podemos neste momento utilizar o Why3 para obter o contra-exemplo seguinte:

\begin{why3}
  let ghost withdraw_stability () : () =
    let ghost accountId = any int in
    (* accountId = 0 *)
    let ghost amount = any int in
    (* amount = 1 *)
    let ghost state = any state in
    assume { amount > 0 /\ 
             ((balances state)[accountId] - amount) >= 0 /\
             accountId >= 0 /\ 
             accountId < length (balances state) };
    withdraw accountId amount state;
    withdraw accountId amount state
    (* accountId = 0; amount = 1 *)
\end{why3}

As afetações produzidas pelo contra-exemplo apresentam-se como comentários. Neste exemplo não é possível garantir que após a execução da primeira chamada a \of{withdraw}, a conta bancária tem um saldo superior a \of{amount}. De acordo com o contra-exemplo, se tentarmos executar a operação \of{withdraw} com \of{amount = 1} sobre a conta \of{accountId = 0} com um saldo inicial de 1, então após a primeira chamada o saldo seria igual a 0. Se executássemos a operação \of{withdraw} novamente com os mesmos argumentos, isso quebraria a pré-condição de \of{withdraw} já que iríamos tentar remover um montante superior ao saldo disponível. Concluímos que não é possível executar duas operações \of{withdraw} em concorrência.

\begin{figure}
    \centering
    \includegraphics[scale=0.75]{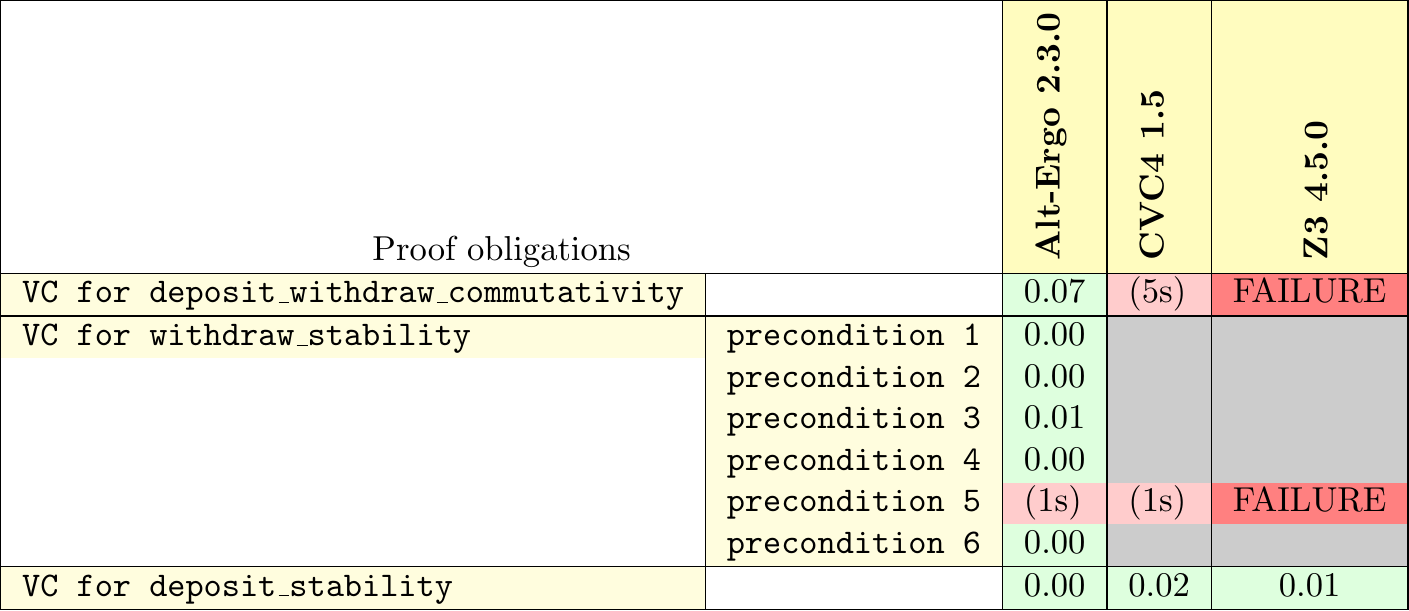}
    \vspace{-5pt}
    \caption{Estatísticas da prova das funções geradas para o caso de estudo.}
    \label{fig:table2}
\end{figure}

O esforço de prova das funções de comutatividade e estabilidade geradas pode ser observado na Figura \ref{fig:table2}.
\vspace{-20pt}

\section{Estado da arte} \label{related_work}

Quelea  é uma ferramenta que verifica aplicações construídas sobre bases de dados replicadas com consistência eventual \cite{quelea}. Esta abordagem tem por base uma linguagem de contratos que permite especificar propriedades \textit{fine-grained} sobre a consistência da aplicação. As condições de verificação derivadas desses contratos são provadas usando o Z3. Para cada operação a ferramenta verifica qual o modelo de consistência apropriado, i.e., cujas restrições são satisfeitas pelo contrato da operação. 
A complexidade da especificação dos contratos é elevada, exigindo racionar sobre possíveis interferências causadas por execuções concorrentes. Para além disso não existe garantia que os contratos sejam suficientes para manter os invariantes da aplicação.

Q9 é  outra ferramenta que analisa aplicações que operam sobre bases de dados replicadas com consistência eventual, dada a sua especificação~\cite{Kaki:2018:SRT:3288538.3276534}. O Q9 descobre anomalias na consistência da aplicação usando uma técnica de verificação limitada, e resolve-as automaticamente. A técnica de verificação limitada consiste em analisar um espaço de procura de estados restringido pelo número de efeitos concorrentes que podem ocorrer ao estado da aplicação. Tendo em conta que existe uma restrição sobre o número máximo de efeitos que podem ocorrer ao estado da aplicação, ao contrário do CISE3, esta ferramenta não é capaz de assegurar correção completa do sistema.

O Repliss é uma ferramenta de verificação de aplicações executadas sobre bases de dados com consistência fraca, dada a especificação, os invariantes e o código da aplicação, escritos na própria \textit{DSL} da ferramenta \cite{Zeller:2017:TPW:3064889.3064893}. Com a informação fornecida pela programadora, o Repliss traduz um programa num programa sequencial Why3. Se o programa sequencial for provado correto, então o programa inicial também é considerado correto. Em comparação com a nossa ferramenta, a \textit{DSL} apresentada nesta ferramenta para a especificação e implementação da aplicação é mais limitada que a linguagem WhyML. Sendo assim, a nossa ferramenta ao ser desenvolvida em sobre a plataforma Why3, permite uma análise mais robusta.

A ferramenta Hamsaz usa  o \textit{SMT solver} CVC4 para determinar 
os conflitos e as dependências causais entre operações~\cite{Hamsaz}, dada um especificação do sistema. Esta especificação deve definir o estado, os invariantes sobre o estado e os métodos da aplicação. O objetivo desta ferramenta é obter automaticamente um sistema replicado correto por construção que garante a integridade e convergência de dados. Em simultâneo, o sistema obtido evita sincronização desnecessária tendo em conta as relações de conflito e dependência entre operações. A análise dos pares de operações em conflito e/ou dependentes é semelhante à análise realizada pela nossa ferramenta.

\section{Conclusão e Perspetivas} 
\label{conclusion}

Neste artigo explorámos uma abordagem automática para a análise de aplicações distribuídas com consistência fraca utilizando a plataforma de verificação dedutiva Why3. O que propomos é que o programa sequencial da aplicação, assim como a sua especificação, sejam fornecidas e de seguida se use a ferramenta CISE3 para raciocinar sobre os pares de operações conflituosas do sistema. A nossa proposta foi ilustrada na Secção \ref{case_study} através dum caso de estudo implementado e verificado na nossa ferramenta. 
Os próximos objetivos relacionados com o trabalho apresentado neste artigo são os seguintes: a especificação e verificação de uma biblioteca de CRDT's, a introdução de uma relação de conflito entre operações no programa e a geração automática de pós-condições mais fortes.

Uma solução para operações que não comutam é a utilização de \textit{Conflict-free Replicated Data Types} (CRDTs) \cite{CRDT_preguica}. Os CRDTs são estruturas de dados, que garantem convergência dos dados, e que estão replicados num conjunto de réplicas ligadas por uma rede assíncrona. 
Com uma biblioteca de CRDTs verificados com a ferramenta Why3, seria possível introduzi-los no programa sempre que fossem identificados pares de operações não comutativas, e desta forma seria possível garantir convergência.

Com a introdução duma relação de conflito no programa, a programadora consegue especificar quais as operações que considera, \textit{à priori} da execução da nossa ferramenta, que estão em conflito. Sendo assim, o nosso objetivo é, dada a relação de conflito entre operações, a nossa ferramenta apenas analisar os pares de operações que não estão representados na relação de conflito especificada pela programadora. Desta forma, a nossa análise apenas gera funções de comutatividade para os pares de operações diferentes que não estejam contemplados na relação de conflito e gera as funções de estabilidade para os pares de operações iguais que não apareçam na relação de conflito.

Com a introdução da geração automática de pós-condições mais fortes, o esforço de especificação do programador seria consideravelmente menor. Neste caso a programadora apenas teria de raciocinar sobre as pré-condições e a implementação de cada operação. Em relação a esta tarefa, o objetivo é endereçar um sub-conjunto da linguagem WhyML de forma a poder obter-se uma prova de conceito para a nossa abordagem.

%
%
%
%

\paragraph{\textbf{Acknowledgements}}
This work was partially supported by NOVA LINCS (UID/CEC/ 04516/2013) and FCT project PTDC/CCI-INF/32081/2017.

\bibliographystyle{splncs04}
\bibliography{bibliography}

\begin{thebibliography}{10}
\providecommand{\url}[1]{\texttt{#1}}
\providecommand{\urlprefix}{URL }
\providecommand{\doi}[1]{https://doi.org/#1}

\bibitem{balegas2015}
Balegas, V., Duarte, S., Ferreira, C., Rodrigues, R., Pregui\c{c}a, N.,
  Najafzadeh, M., Shapiro, M.: Putting consistency back into eventual
  consistency. In: EuroSys (2015)

\bibitem{balegas2016}
Balegas, V., Li, C., Najafzadeh, M., Porto, D., Clement, A., Duarte, S.,
  Ferreira, C., Gehrke, J., Leit{\~a}o, J., Pregui{\c c}a, N., Rodrigues, R.,
  Shapiro, M., Vafeiadis, V.: {Geo-Replication: Fast If Possible, Consistent If
  Necessary}. {IEEE Data Engineering Bulletin}  \textbf{39}(1) (2016)

\bibitem{boogie}
Barnett, M., Chang, B.Y.E., DeLine, R., Jacobs, B., Leino, K.R.M.: Boogie: A
  modular reusable verifier for object-oriented programs. In: FMCO (2005)

\bibitem{DBLP:journals/jlp/DaillerHMM18}
Dailler, S., Hauzar, D., March{\'{e}}, C., Moy, Y.: Instrumenting a weakest
  precondition calculus for counterexample generation. J. Log. Algebr. Meth.
  Program.  \textbf{99} (2018)

\bibitem{dailler2018}
Dailler, S., March{\'e}, C., Moy, Y.: Lightweight interactive proving inside an
  automatic program verifier. In: F-IDE (2018)

\bibitem{memoire}
Filli{\^a}tre, J.C.: {Deductive Software Verification}. International Journal
  on Software Tools for Technology Transfer  \textbf{13}(5) (2011)

\bibitem{ghost_code}
Filli{\^a}tre, J.C., Gondelman, L., Paskevich, A.: {The Spirit of Ghost Code}.
  In: {CAV} (2014)

\bibitem{krakatoa}
Filli{\^a}tre, J.C., March{\'e}, C.: The why/krakatoa/caduceus platform for
  deductive program verification. In: CAV (2017)

\bibitem{filliatre:hal-00789533}
Filli{\^a}tre, J.C., Paskevich, A.: {Why3 -- Where Programs Meet Provers}. In:
  {ESOP} (2013)

\bibitem{modular_way}
Filli{\^a}tre, J.C., Pereira, M.: {A Modular Way to Reason About Iteration}.
  In: NFM (2016)

\bibitem{CAP}
Gilbert, S., Lynch, N.: Brewer's conjecture and the feasibility of consistent,
  available, partition-tolerant web services. SIGACT News  \textbf{33}(2)
  (2002)

\bibitem{cise}
Gotsman, A., Yang, H., Ferreira, C., Najafzadeh, M., Shapiro, M.: {'Cause {I}'m
  Strong Enough: Reasoning About Consistency Choices in Distributed Systems}.
  In: POPL (2016)

\bibitem{spark}
Hoang, D., Moy, Y., Wallenburg, A., Chapman, R.: Spark 2014 and gnatprove.
  International Journal on Software Tools for Technology Transfer
  \textbf{17}(6) (2015)

\bibitem{Hamsaz}
Houshmand, F., Lesani, M.: Hamsaz: Replication coordination analysis and
  synthesis. In: POPL (2019)

\bibitem{Kaki:2018:SRT:3288538.3276534}
Kaki, G., Earanky, K., Sivaramakrishnan, K., Jagannathan, S.: Safe replication
  through bounded concurrency verification. In: OOPSLA (2018)

\bibitem{Kirchner:2015:FSA:2769048.2769092}
Kirchner, F., Kosmatov, N., Prevosto, V., Signoles, J., Yakobowski, B.:
  Frama-c: A software analysis perspective. Form. Asp. Comput.  \textbf{27}(3)
  (2015)

\bibitem{Marcelino:2017:BHC:3064889.3064896}
Marcelino, G., Balegas, V., Ferreira, C.: Bringing hybrid consistency closer to
  programmers. In: PaPoC (2017)

\bibitem{jessie}
Marh{\'e}, C., Moy, Y.: The jessie plugin for deductive verification in
  frama-c. INRIA Saclay {\^I}le-de-France and LRI, CNRS UMR  (2012)

\bibitem{nair:hal-01832888}
Nair, S.S., Shapiro, M.: {Improving the ''Correct Eventual Consistency'' Tool}.
  Research Report RR-9191, {Sorbonne Universit{\'e}} (2018)

\bibitem{Najafzadeh:2016:CTP:2911151.2911160}
Najafzadeh, M., Gotsman, A., Yang, H., Ferreira, C., Shapiro, M.: {The CISE
  Tool: Proving Weakly-consistent Applications Correct}. In: PaPoC (2016)

\bibitem{CRDT_preguica}
Shapiro, M., Pregui{\c{c}}a, N., Baquero, C., Zawirski, M.: Conflict-free
  replicated data types. In: SSS (2011)

\bibitem{quelea}
Sivaramakrishnan, K., Kaki, G., Jagannathan, S.: Declarative programming over
  eventually consistent data stores. In: PLDI (2015)

\bibitem{Zeller:2017:TPW:3064889.3064893}
Zeller, P.: Testing properties of weakly consistent programs with repliss. In:
  PaPoC (2017)

\end{thebibliography}

\appendix

%
%

\end{document}